\documentclass[reprint, superscriptaddress, amsmath, amssymb, prl]{revtex4-2}
\usepackage{graphicx,dcolumn,xcolor,blindtext,mathptmx,etoolbox,bm,float}
\usepackage[utf8]{inputenc}
\usepackage[T1]{fontenc}
\usepackage[normalem]{ulem}
\definecolor{prlblue}{RGB}{46, 48, 146}
\usepackage[colorlinks=true,citecolor=prlblue,urlcolor=prlblue,linkcolor=black]{hyperref}

\begin{document}

\title{Atomically smooth films of CsSb: a chemically robust visible light photocathode}

\author{C. T. Parzyck}
  \affiliation{Laboratory of Atomic and Solid State Physics, Department of Physics, Cornell University, Ithaca, NY 14853, USA}
\author{C. A. Pennington}
  \affiliation{Cornell Laboratory for Accelerator-Based Sciences and Education, Cornell University, Ithaca, NY 14853, USA}
\author{W. J. I. DeBenedetti}
  \affiliation{Department of Chemistry and Chemical Biology, Cornell University, Ithaca, NY 14853, USA}
\author{J. Balajka}
  \affiliation{Department of Chemistry and Chemical Biology, Cornell University, Ithaca, NY 14853, USA}
\author{E. Echeverria}
  \affiliation{Cornell Laboratory for Accelerator-Based Sciences and Education, Cornell University, Ithaca, NY 14853, USA}

\author{H. Paik}
  \affiliation{Platform for the Accelerated Realization, Analysis, and Discovery of Interface Materials (PARADIM), Cornell University, Ithaca, NY 14853, USA}
  \affiliation{School of Electrical \& Computer Engineering, University of Oklahoma, Norman, Oklahoma 73019}
  \affiliation{Center for Quantum Research and Technology, University of Oklahoma, Norman, Oklahoma 73019}
\author{L. Moreschini}
\affiliation{Platform for the Accelerated Realization, Analysis, and Discovery of Interface Materials (PARADIM), Cornell University, Ithaca, NY 14853, USA}
\author{B. D. Faeth}
 \affiliation{Platform for the Accelerated Realization, Analysis, and Discovery of Interface Materials (PARADIM), Cornell University, Ithaca, NY 14853, USA}
\author{C. Hu}
  \affiliation{Platform for the Accelerated Realization, Analysis, and Discovery of Interface Materials (PARADIM), Cornell University, Ithaca, NY 14853, USA}
\author{J. K. Nangoi}
  \affiliation{Laboratory of Atomic and Solid State Physics, Department of Physics, Cornell University, Ithaca, NY 14853, USA}
\author{V. Anil}
  \affiliation{Laboratory of Atomic and Solid State Physics, Department of Physics, Cornell University, Ithaca, NY 14853, USA}

\author{T. A. Arias}
  \affiliation{Laboratory of Atomic and Solid State Physics, Department of Physics, Cornell University, Ithaca, NY 14853, USA}
\author{M. A. Hines}
  \affiliation{Department of Chemistry and Chemical Biology, Cornell University, Ithaca, NY 14853, USA}
\author{D. G. Schlom}
  \affiliation{Department of Materials Science and Engineering, Cornell University, Ithaca, NY 14853, USA}
  \affiliation{Kavli Institute at Cornell for Nanoscale Science, Cornell University, Ithaca, NY 14853, USA}
  \affiliation{Leibniz-Institut f{\"u}r Kristallz{\"u}chtung, Max-Born-Stra{\ss}e 2, 12489 Berlin, Germany}
\author{A. Galdi}
  \affiliation{Cornell Laboratory for Accelerator-Based Sciences and Education, Cornell University, Ithaca, NY 14853, USA}
  \affiliation{Department of Industrial Engineering, University of Salerno, Fisciano (SA) 84084, Italy}
\author{K. M. Shen}
  \affiliation{Laboratory of Atomic and Solid State Physics, Department of Physics, Cornell University, Ithaca, NY 14853, USA}
  \affiliation{Kavli Institute at Cornell for Nanoscale Science, Cornell University, Ithaca, NY 14853, USA}
\author{J. M. Maxson}
  \affiliation{Cornell Laboratory for Accelerator-Based Sciences and Education, Cornell University, Ithaca, NY 14853, USA}
  \email[Author to whom correspondence should be addressed: ]{agaldi@unisa.it; jmm586@cornell.edu}

\date{\today}

\begin{abstract}
 Alkali antimonide semiconductor photocathodes provide a promising platform for the generation of high brightness electron beams, which are necessary for the development of cutting-edge probes including x-ray free electron lasers and ultrafast electron diffraction. However, to harness the intrinsic brightness limits in these compounds, extrinsic degrading factors, including surface roughness and contamination, must be overcome.  By exploring the growth of Cs$_x$Sb thin films monitored by \textit{in situ} electron diffraction, the conditions to reproducibly synthesize atomically smooth films of CsSb on 3C-SiC (100) and graphene coated TiO$_2$ (110) substrates are identified, and detailed structural, morphological, and electronic characterization is presented. These films combine high quantum efficiency in the visible (up to 1.2\% at 400 nm), an easily accessible photoemission threshold of 550 nm, low surface roughness (down to 600 pm on a 1 $\mu$m scale), and a robustness against oxidation up to 15 times greater then Cs$_3$Sb. These properties suggest that CsSb has the potential to operate as an alternative to Cs$_3$Sb in electron source applications where the demands of the vacuum environment might otherwise preclude the use of traditional alkali antimonides.\\
 
 keywords: photocathodes, molecular beam epitaxy, surface science, electronic structure.
\end{abstract}
\maketitle
%%%%%%%%%%%%%%%%%%%%%%%%%%%%%%%%% Introduction %%%%%%%%%%%%%%%%%%%%%%%%%%%%%%%%%
%%%%%%%%%%%%%%%%%%%%%%%%%%%%%%%%%%%%%%%%%%%%%%%%%%%%%%%%%%%%%%%%%%%%%%%%%%%%%%%%
\section{Introduction}
High brightness electron beams are an essential ingredient in a variety modern scientific applications which require high charge and ultrashort electron pulses. These applications range from x-ray free electron lasers (FEL), \cite{euxfel,LCLS} to fs-scale ultrafast electron microscopes, \cite{4dem, dwayne, medusa} and to electron-based hadron cooling systems and electron linear colliders.\cite{ilc, cec, ccubed} Generation of pulsed electron beams is accomplished via photoemission of electrons from specifically tailored materials characterized by high quantum efficiency (QE, photoemitted electrons per incident photon).  However, to ensure high brightness of the resulting beam, the intrinsic emittance of the material (a measurement of the momentum spread of the photoelectrons) must be minimized \cite{tradeoff} by limiting the physical and chemical roughness of the material surface. \cite{Gevorkian}  One class of materials identified by accelerator scientists as high efficiency photocathode candidates are alkali antimonide semiconductors: AA$'_{2}$Sb (A,A'=Cs,K,Na,Rb, including A=A'). \cite{Lewellen:2022oeb} These compounds are characterized by QE's of $10^{-2}$ to $10^{-1}$ at $\sim550$~nm \cite{sommer} and by mean transverse energies below 180~meV at 532~nm, which can be further reduced by operating near the photoemission threshold and at low temperature. These low MTEs are relevant, for example, for next generation high repetition rate FELs. \cite{Lewellen:2022oeb, LucaCold} However, this class of materials is also extremely sensitive to oxidation -- which tends to suppress the photocathode efficiency and enlarge the MTE.  As such, alkali antimonides have stringent vacuum requirements, demanding pressures below $10^{-10}$ Torr to be handled without significant degradation,\cite{Bates1981, Hines} which limits the scope of their applicability. Additionally, the high vapor pressure of alkali metals at ambient temperatures presents a challenge to the synthesis of smooth, ordered films. \cite{Saha:2022glf, SahaAPL, GaldiIPAC, Epitaxy} Because surface disorder induces emittance degradation and reduces the utility of the photocathode, significant effort has been devoted to finding ways to reduce the crystalline disorder of these materials.  Recent advances include both improving the as-grown film properties (smoothness, homogeneity) though new synthesis techniques\cite{Feng-rough, roughnessGaAs, Galdi_roughness} as well as finding ways to increase their robustness against contamination and ageing -- for example by encapsulating them in 2D materials. \cite{armor, Biswas_encapsulated}

In this article we explore the phase diagram of Cs$_x$Sb using a variety of \textit{in operando} and \textit{in situ} probes of the structure, morphology, and photoemission properties of the resulting films. We identify another member of the alkali antimonide family, CsSb, as a visible light photoemitter which can be grown in crystalline, ultra-flat films characterized by surface roughness better than 1 nm on a 1 $\mu$m scale and quantum efficiency up to 1.2\% at 400~nm. Near-threshold operation of AA$'_{2}$Sb photocathodes requires the use of $\sim650-700$~nm light to achieve low emittance, \cite{LucaCold} requiring complex optical schemes (\textit{e.g.} optical parametric amplification) to reach. On the other hand, the photoemission threshold of CsSb is found to be close to 550~nm, which means that near-threshold operation can use the second harmonic of many common high repetition rate laser gain media, implying significant relief on the optical components and on the light source power. Finally, this phase is found to be substantially more resilient to oxygen contamination than its more commonly synthesized cousin, Cs$_3$Sb, which may extend the operational lifetime of the cathodes in the demanding environment of a photoinjector cavity.

The recent achievement of epitaxial single-oriented Cs$_3$Sb films\cite{Epitaxy} was in large part made possible through the use of reflection high energy electron diffraction (RHEED) as a structural diagnostic during growth.  This technique is a mainstay of traditional semiconductor, metal, and oxide film growth by molecular beam epitaxy (MBE); however, to date it has been infrequently employed in photocathode preparation, where the traditional \textit{in operando} diagnostic is the QE. \cite{GaldiNAPAC, Pavlenko} By monitoring the structure of the sample during deposition, using RHEED, we identify various growth regimes of Cs$_x$Sb as a function of the growth temperature, including the stabilization of ordered films of CsSb.  We study the chemical composition of the resulting films using x-ray photoemission spectroscopy (XPS), their morphology using scanning tunneling microscopy (STM), and their electronic structure using angle resolved photoemission spectroscopy (ARPES).

%%%%%%%%%%%%%%%%%%%%%%%%%%%%%%%% Sample Growth %%%%%%%%%%%%%%%%%%%%%%%%%%%%%%%%%
%%%%%%%%%%%%%%%%%%%%%%%%%%%%%%%%%%%%%%%%%%%%%%%%%%%%%%%%%%%%%%%%%%%%%%%%%%%%%%%%
\section{Synthesis}
\begin{figure}
  \includegraphics[width=8.5cm]{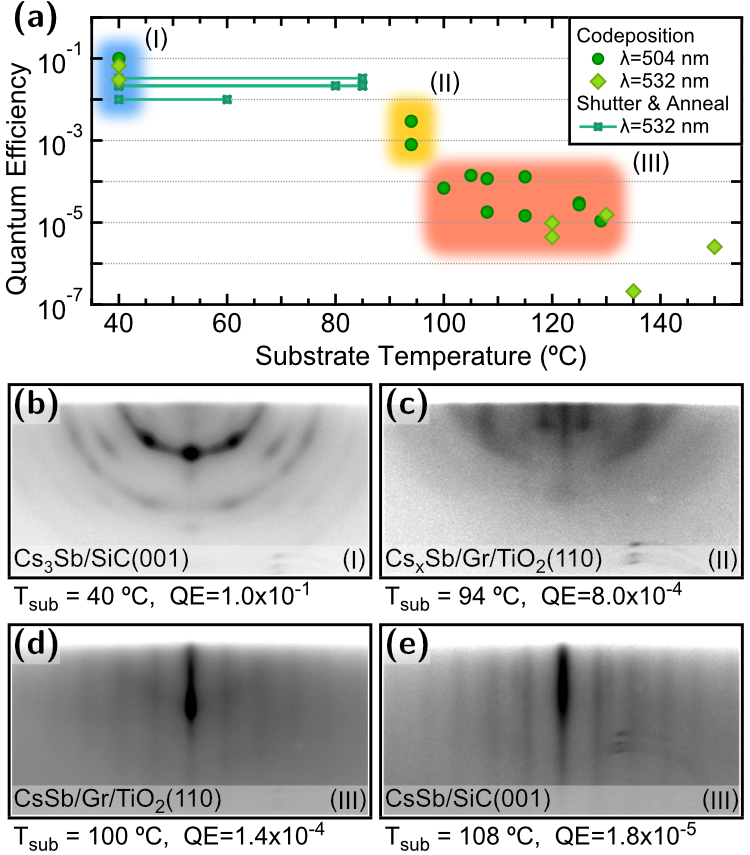}
  \caption{\label{fig:Growth} Quantum efficiency (QE) of Cs$_x$Sb photoemitters, $1\leq x\leq 3$. (a) QE (in the green) as a function of growth temperature.  Samples grown by codeposition at a single temperature are shown as circles and diamonds; samples using the solid-phase epitaxy approach of Ref. \onlinecite{Epitaxy} are depicted as lines connecting the deposition and annealing temperatures. (b) RHEED pattern of a textured, high-efficiency Cs$_3$Sb film grown at low temperature.  (c) RHEED pattern of a textured, medium-efficiency film in the intermediate regime.  (d)-(e) RHEED patterns of fiber-texture CsSb films codeposited at temperatures above 100 $^{\circ}$C.  Panels (b)-(e) use a logarithmic intensity scale, quoted QEs are measured at 504 nm.}
\end{figure}

We begin with an overview of the growth of Cs$_x$Sb photocathodes over a range of conditions with the structure monitored during growth using RHEED and the QE measured directly afterwards.  Because the cesium desorption rate from the sample is strongly temperature dependent in the range explored here, adsorption control of the stoichiometry can be accomplished by oversupplying cesium and varying the substrate temperature, rather then by varying the Cs:Sb flux ratio.  The quantum efficiency of a set of samples grown under similar flux conditions (Cs:Sb between 6.0 and 6.6) are depicted in \hyperref[fig:Growth]{Figure \ref*{fig:Growth}(a)}.  Using RHEED, we identify three distinct regimes of film growth across the explored temperature range.  At low temperatures, regime (I) -- around 40~$^{\circ}$C, the high efficiency Cs$_3$Sb phase is formed with QE ranging from 3 to 10\%.  When films are codeposited at this temperature they are polycrystalline and form textured ring patterns in RHEED, shown in \hyperref[fig:Growth]{Figure \ref*{fig:Growth}(b)}.  Once formed, however, this phase is stable against Cs loss at higher temperatures and can be annealed up to $\sim 85$~$^{\circ}$C to order the domains and, on an appropriate substrate, produce an epitaxial film\cite{Epitaxy} while maintaining ~1\% level QE in the green. When films are depositied at higher temperatures, $\sim 90$~$^{\circ}$C -- regime (II), the RHEED patterns show less well defined rings and some faint streaks.  While the QE of these samples remains reasonably high ($\sim 10^{-3}$ at 504 nm) RHEED indicates degraded crystallinity and a lack of ordering.  However, when the substrate temperature is further increased, exceeding 100~$^{\circ}$C -- regime (III), a new phase emerges which is characterized by the streaked RHEED pattern shown in \hyperref[fig:Growth]{Figure \ref*{fig:Growth}(d-e)}.  As will be subsequently discussed, spectroscopic measurements identify the stoichiometry of this phase to be CsSb.  No azimuthal dependence of the RHEED streaks is observed, indicating no preferential in-plane orientation of the films. Nonetheless, the absence of rings indicates alignment of the out-of-plane axes of the grains and the lack of vertical modulation indicates smooth, terraced growth of a so-called `fiber textured' film.\cite{Tang_2007}  Though the QE of films grown in this regime is reduced somewhat (ranging from $1.1\times10^{-5}$ to $1.4\times10^{-4}$ at 504 nm) compared to Cs$_3$Sb, these films remain visible light photoemitters.  The growth window for this particular phase appears about 30~$^{\circ}$C wide, with the quantum efficiency dropping below $10^{-5}$ at higher growth temperatures.  The remainder of this work is concerned with the study of films in regime (II) and (III), including their morphology and photoemission properties.  We note that in regime (III) fiber-textured films are produced on both of the substrates investigated here, 3C-SiC (100) and monolayer graphene deposited on rutile TiO$_2$ (110).  No discernible differences in either the RHEED patterns or QE were observed between films grown on the two substrates, indicating they both provide reasonable platforms for the synthesis of CsSb.

%%%%%%%%%%%%%%%%%%%%%%%%%%%%%%%%%%%% XPS %%%%%%%%%%%%%%%%%%%%%%%%%%%%%%%%%%%%%%%
%%%%%%%%%%%%%%%%%%%%%%%%%%%%%%%%%%%%%%%%%%%%%%%%%%%%%%%%%%%%%%%%%%%%%%%%%%%%%%%%
\section{X-ray Photoelectron Spectroscopy}
\begin{figure*}
  \includegraphics[width=17 cm]{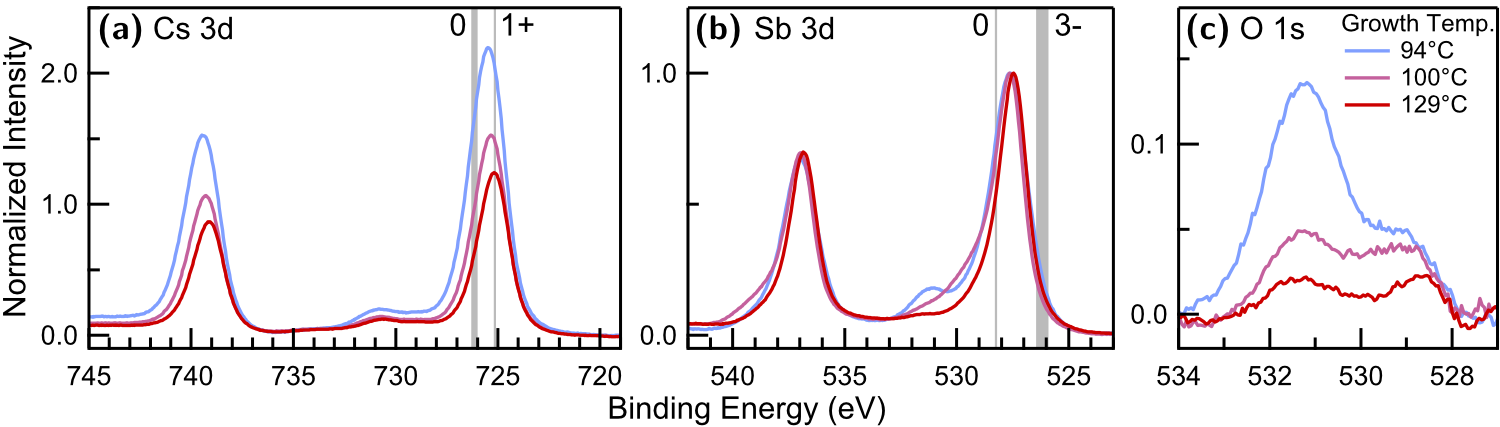}
  \caption{\label{fig:XPS}  Background subtracted XPS spectra in the (a) Cs~3d, (b) Sb~3d and (c) O~1s regions for samples grown at different substrate temperatures; intensity is normalized to Sb 3d$_{5/2}$ peak.  Literature reference energies for Cs$^0$, Cs$^{1+}$, and Sb$^{3-}$ are provided as grey lines\cite{Schmeisser,EBBINGHAUS,Krishnan_1977} in addition to a metallic Sb sample (Sb$^0$) measured in the same XPS system. \cite{Epitaxy}  Measurements presented here were performed at a grazing emission angle of 70$^\circ$, normal emission data is available in the supplemental materials.}
\end{figure*}
In \hyperref[fig:XPS]{Figure \ref*{fig:XPS}(a-b)} we report the Cs~3d and Sb~3d XPS spectra of three Cs$_x$Sb samples grown on graphene/TiO$_2$ (110) at different substrate temperatures ($T_{sub}$), one deposited in growth regime (II) ($T_{sub}=94^\circ$C) and two grown at higher temperature, in regime (III). In \hyperref[fig:XPS]{Figure \ref*{fig:XPS}(a)}, the Cs~3d$_{5/2}$ peak position is close to the Cs$^+$ reference energy (752.2~eV) for all the samples, though some shift towards higher binding energies is observed with decreasing T$_{sub}$. The peak at $\sim 730.9$~eV is attributable to the Mg K$\alpha_3$ satellite of the Cs~3d$_{3/2}$ peak, although the intensity ratio between the peaks, exceeding 8\%, does not exclude some plasmon contribution.  In contrast to measurements of high QE Cs$_3$Sb\cite{MARTINI2015} and metallic Cs, \cite{XPSbook} no strong plasmon peaks are observed in these samples.  The Sb~3d$_{5/2}$ peak position, 527.45-527.65 eV, falls between the binding energies of Sb$^0$ metal (528.3~eV) and Sb$^{3-}$ in Cs$_3$Sb (526.1~eV) reference samples (a comparison is reported in supplementary materials). In previous studies of Cs-Sb compounds, this binding energy value has been attributed to Sb$^0$ (either bulk or ``atomic"), \cite{Soriano1993, Cocchi} while others attributed it to different phases of the Cs-Sb system, such as CsSb or Cs$_5$Sb$_4$. \cite{Bates1981} However, evaluation of the Sb Auger parameter in the present films (supplementary materials) is consistent with reduced Sb in Cs$_3$Sb rather than Sb metal. Furthermore, estimates of the sample composition, reported in \hyperref[composition]{Table \ref*{composition}}, are close to Cs:Sb=1:1 which leads us to attribute the observed Sb~3d$_{5/2}$ binding energy value to Sb$^{1-}$ species. 

To enhance surface sensitivity and examine the presence of surface oxidation, the spectra of \hyperref[fig:XPS]{Figure \ref*{fig:XPS}} were collected in grazing emission. \cite{Hines}.  Though the O~1s spectrum ([528, 534]~eV) overlaps with the Sb~3d peaks, the spin-orbit splitting of the Sb~3d$_{3/2}$-3d$_{5/2}$ peaks can be exploited to isolate the oxygen contribution using the methods of Ref. \onlinecite{Hines}; the results are shown in \hyperref[fig:XPS]{Figure \ref*{fig:XPS}(c)}. In all three samples we observe some contribution from oxygen species, most probably originating from exposure during the vacuum suitcase sample transfer between the MBE and STM-XPS systems. \cite{Epitaxy} We observe O~1s binding energies of 531.2~eV and 528.7-529~eV, the latter close to the expected Sb 3d$_{3/2}$ satellite ($\sim528.4$~eV). Both energies fall into the range associated to metal oxides, \cite{XPSbook} and in particular the former is close to the binding energy associated to peroxide species in Cs$_2$O$_2$ (529.9-531.0 eV)\cite{JUPILLE}, or to antimony suboxide. \cite{Liu}

The surface composition of the samples of \hyperref[fig:XPS]{Figure \ref*{fig:XPS}} was obtained from the integrated intensity of Cs 3d, Sb 3d and O 1s spectra normalized by their relative sensitivity factors and photoelectron escape depth; the results are reported in Table~\ref{composition}. From the less surface-sensitive normal emission measurements, we observe that the Cs:Sb ratio is closest to 1:1 for the sample grown at higher temperature, and that Cs content increases with decreasing growth temperature. Since all pictured samples were stored and transferred together from the MBE to the STM-XPS system they received identical exposure to residual gasses.  Therefore, the observed correlation between lower oxygen content and reduced Cs:Sb ratio in samples grown at higher $T_{sub}$ indicates that stoichiometric CsSb has an increased oxidation resistance over the Cs rich phases. The Cs and O content is higher in the more surface sensitive measurements at grazing emission, which is consistent with surface oxidation. However, the differences between the spectra at different emission angles (supplementary materials) indicates that the compositional gradient through the samples is minor, especially in comparison to the typical behaviour of Cs$_3$Sb where Cs segregation is observed in response to any oxygen exposure. \cite{Hines, GaldiNAPAC}  Using the methods described in ref. \onlinecite{Morant, Soriano1993}, the XPS data can be modeled by a bottom layer with composition Cs:Sb~$\approx1:1$ covered by a surface layer with composition Cs:O~$\approx1:1$, consistent with a layer of Cs$_2$O$_2$, although the model parameters cannot be unequivocally determined using only two emission angles. The observed shift of the Cs~3d binding energy and the less pronounced one of the Sb~3d can be explained by band-bending induced by Cs$_2$O$_2$, \cite{Sun} analogous to observations of superficially oxidized Cs$_3$Sb. \cite{Hines}

\begin{table}[htb]
 \begin{ruledtabular}
 \caption{\label{composition} Composition of CsSb samples obtained by XPS as a function of growth temperature, $T_{sub}$, and photoelectron emission angle: normal emission = 0$^\circ$, grazing emission = 70$^\circ$). At grazing emission the probe depth is reduced by a factor of $\sim 3$, enhancing sensitivity to the sample surface.}
 \begin{tabular}{ccc}
$T_{sub}$~($^\circ$C) & 0$^\circ$ &  70$^\circ$ \\
 \hline
 94 & Cs$_{0.51}$O$_{0.12}$Sb$_{0.37}$&Cs$_{0.53}$O$_{0.21}$Sb$_{0.26}$\\
  100 & Cs$_{0.48}$O$_{0.11}$Sb$_{0.41}$&Cs$_{0.51}$O$_{0.13}$Sb$_{0.36}$\\
  129 & Cs$_{0.48}$O$_{0.07}$Sb$_{0.45}$&Cs$_{0.49}$O$_{0.09}$Sb$_{0.41}$\\
 \end{tabular}
 \end{ruledtabular}
 \end{table}

%%%%%%%%%%%%%%%%%%%%%%%%%%%%%%%%%%%% STM %%%%%%%%%%%%%%%%%%%%%%%%%%%%%%%%%%%%%%%
%%%%%%%%%%%%%%%%%%%%%%%%%%%%%%%%%%%%%%%%%%%%%%%%%%%%%%%%%%%%%%%%%%%%%%%%%%%%%%%%
\section{Scanning tunneling microscopy}
\begin{figure}
  \includegraphics[width=8.5cm]{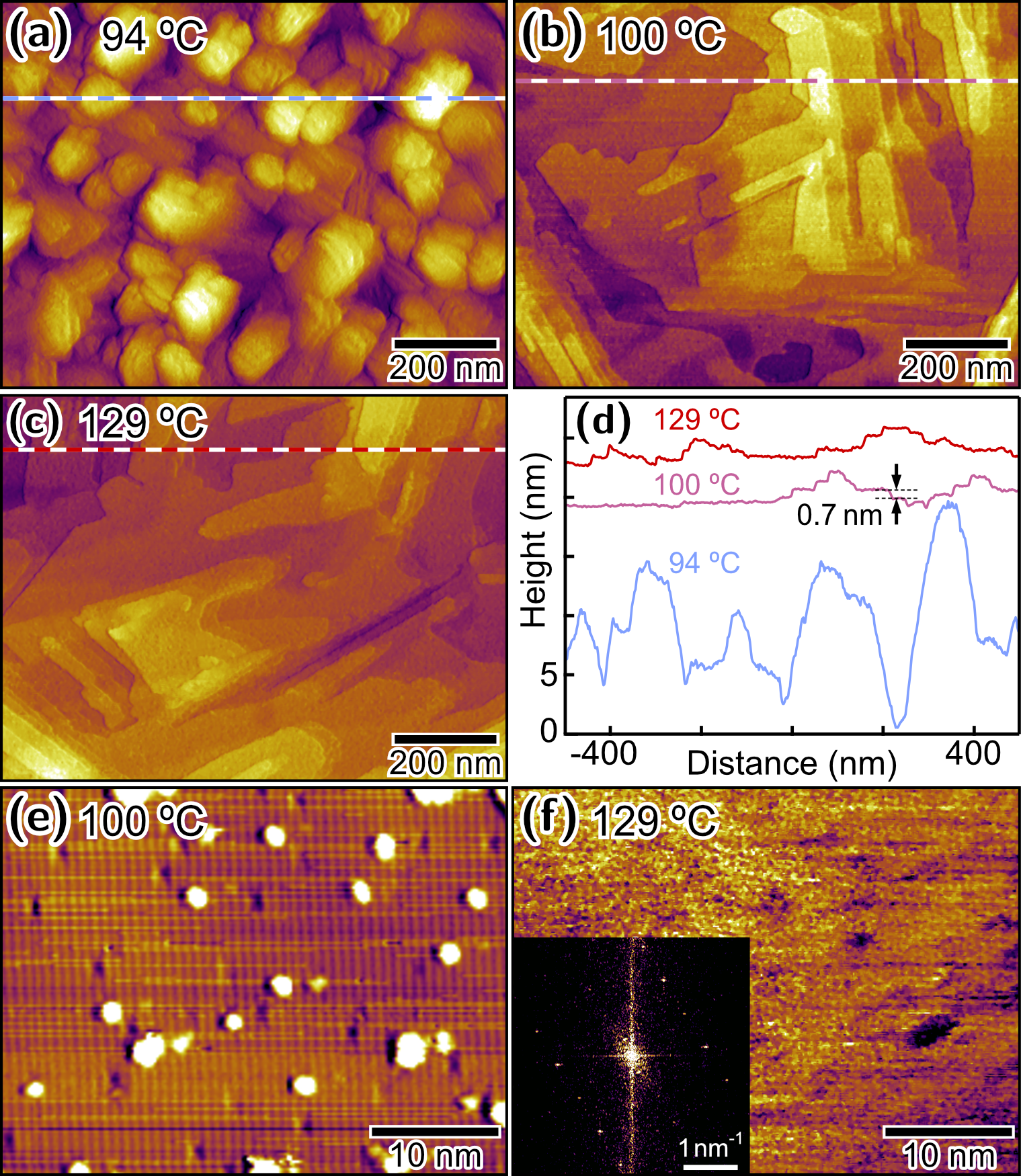}
  \caption{\label{fig:STM} STM images of Cs$_x$Sb films grown on Graphene coated TiO$_2$ (110) substrates. (a-c) $1.0\mu m \times 0.76\mu m$ images of samples grown at 94 $^\circ$C, 100 $^\circ$C and 129 $^\circ$C, respectively. (d) Line cuts from the marked areas in the prior three images, traces are offset for clarity.  (e),(f) higher magnification images of terraces imaged in samples in (b) and (c) respectively; inset shows the Fourier transform of the STM height map.}
\end{figure}

  A major stumbling block in the development of alkali antimonide photocathodes has been their propensity to form rough or disordered surfaces during growth, contributing to high MTEs of the photoemitted electrons.  Recent advances, including careful choice of substrate and growth temperature, have helped to mitigate the surface roughness of Cs$_3$Sb films\cite{Galdi_roughness,SahaAPL,Epitaxy}, but the synthesis of atomically ordered films still requires a delicate multi-step shuttered growth procedure.   One advantage of CsSb hinted at by the RHEED patterns in \hyperref[fig:Growth]{Figures \ref*{fig:Growth}(d-e)} is that smooth, terraced films may be produced by single temperature codeposition.  To quantify this, the morphology of a set of MBE grown CsSb films (on monolayer graphene coated TiO$_2$ (110)) was investigated using STM, and the results are summarized in \hyperref[fig:STM]{Figure \ref*{fig:STM}} and the supplemental materials.  A sample grown in regime (II) at a lower temperature of $\sim 94^\circ$C shows a rough and disordered grain structure with a characteristic grain size of $\sim85$~nm (equivalent disc radius) and a root-mean-square (rms) roughness of 1.4 nm averaged over grains.  The surface roughness averaged over a 1$\mu$m x 1$\mu$m area is found to be 2.3 nm.  However, samples grown at higher temperatures, in regime (III) show flat, smooth terraces, exampled in \hyperref[fig:STM]{Figures \ref*{fig:STM}(b-c)}.  The terrace-averaged rms roughness observed in \hyperref[fig:STM]{Figure \ref*{fig:STM}(b)} is only 240 pm (averaged over a lateral scale of $\sim 200$~nm) and is 600 pm when averaged on the 1~$\mu$m scale.  The sample grown at higher temperature, in \hyperref[fig:STM]{Figure \ref*{fig:STM}(c)}, shows a slightly increased rms roughness of 750 pm on the 1~$\mu$m scale.  These roughness values compare favorably to state-of-the-art codeposited films of Cs$_3$Sb on SiC \cite{Galdi_roughness} and SrTiO$_3$. \cite{Saha2022}

\hyperref[fig:STM]{Figure \ref*{fig:STM}(d)} shows a set of representative lineouts from the previous three images illustrating the terrace widths and heights.  The terraces are separated by steps of roughly 0.7 nm, though there is some variation in the step heights measured across the STM maps, and steps between 0.5 and 0.9 nm are observed.  A higher magnification image of one of the terraces is given in \hyperref[fig:STM]{Figure \ref*{fig:STM}(e)} which shows atomic columns decorated with both pits and islands, likely due to the presence of both atom vacancies and adatoms on the sample surface. The column-to-column distance in this image is 0.79 nm, and the atomic columns change orientation at grain boundaries (\textit{c.f.} supplementary materials) as expected from the fiber-texture diffraction pattern observed in RHEED.  Together, the RHEED and STM indicate that the films grow locally in ordered crystalline domains, with length scales of 100-200 nm, which are rotationally misaligned to form an even distribution on the macroscale.  As is shown in \hyperref[fig:STM]{Figure \ref*{fig:STM}(c)} the terraced structure of the films is preserved throughout regime (III), and while atomic rows were not visible in the higher magnification imaging of the terraces, \hyperref[fig:STM]{Figure \ref*{fig:STM}(f)}, a similar in-plane lattice constant of 0.76-0.78 nm can be extracted from the Fourier transform of the height map. 

\begin{figure}
  \includegraphics[width=8.49cm]{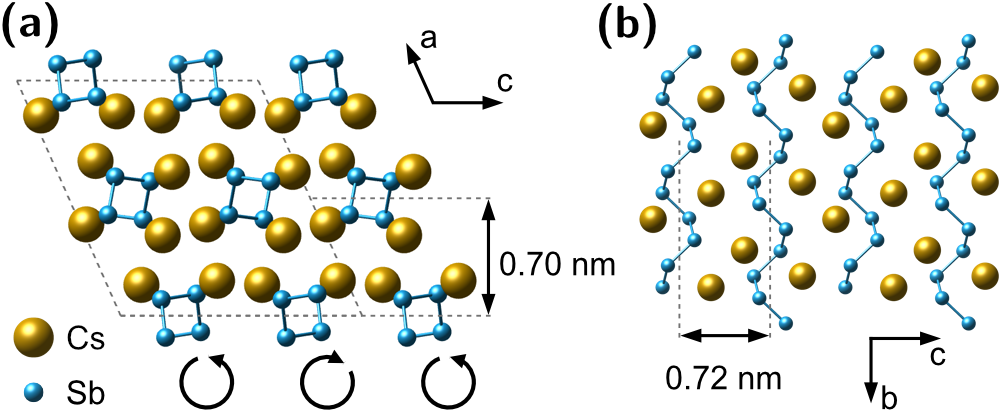}
  \caption{\label{fig:atoms} Visualization of the crystal structure\cite{vesta} of the monoclinic phase, $\beta$--CsSb, measured in Ref. \onlinecite{Los2002}.  (a) Projection along [010] showing the layered structure of the Sb spirals.  Alternating layers have opposing tilts and alternating columns have alternating chirality. (b) Structure of a single $\beta$--CsSb monolayer, viewed down the [100] axis, illustrating the quasi-1D structure of the Sb spirals.}
\end{figure}

We turn now to a discussion of the crystal structure of our films, based on the morphological evidence from STM.  It has been shown previously that bulk CsSb may crystallize in one of two related structures.  It was originally discovered that at higher temperatures ($>500~^{\circ}$C) and longer reaction times ($>100$~hours) a NaP type orthorhombic phase with space group $P2_12_12_1$ forms \cite{Honle1979}.  However it was later observed that at lower temperatures ($\sim440~^{\circ}$C) and shorter reaction times ($\sim$1 hour) a monoclinic phase may form with space group $P2_1/c$ \cite{Los2002}.  Following Emmerling et al. we term the high temperature orthorhombic phase as $\alpha-$CsSb and the lower temperature monoclinic phase as $\beta-$CsSb.  Though the crystal symmetries differ, the $\alpha$ and $\beta$ phases share a common structural motif, being composed of extended chains of Sb atoms surrounded in a cage of Cs ions -- it is the relative orientation of the chains and their stacking sequence which differentiates the two phases.  A visualization of the $\beta$ phase based on crystallography data reported in Ref. \onlinecite{Los2002} is provided in \hyperref[fig:atoms]{Figure \ref*{fig:atoms}} and discussion of the $\alpha$ phase is provided in the supplementary materials. For consistency, in both structures we take the axis along the chains to be the [010] direction.  In the $\beta$ phase the chirality of the Sb chains alternates across the (001) planes and the relative rotation of the chains alternates across (100).  The $\beta$ phase is composed of monolayers of chains stacked along the [100] direction with an expected spacing of 0.70 nm, which is also consistent with the measured STM step heights. Additionally, the chain-to-chain spacing expected from bulk measurements for (100) oriented $\beta-$CsSb is 0.72 nm, which is relatively consistent with the above STM measurements.  Note that presence of the $\alpha$ phase cannot be ruled out entirely by the measurements performed here as determination of the crystal symmetry by \textit{ex situ} x-ray diffraction was not possible due to the air-sensitivity of the samples.  Further refinement of the structure using \textit{in situ} x-ray diffraction would help to clarify the precise phase stabilized under these growth conditions.

%%%%%%%%%%%%%%%%%%%%%%%%%%%%%%%%% UPS/ARPES %%%%%%%%%%%%%%%%%%%%%%%%%%%%%%%%%%%%
%%%%%%%%%%%%%%%%%%%%%%%%%%%%%%%%%%%%%%%%%%%%%%%%%%%%%%%%%%%%%%%%%%%%%%%%%%%%%%%%
\section{Angle-resolved Photoemission Spectroscopy}
\begin{figure*}
  \includegraphics[width=17cm]{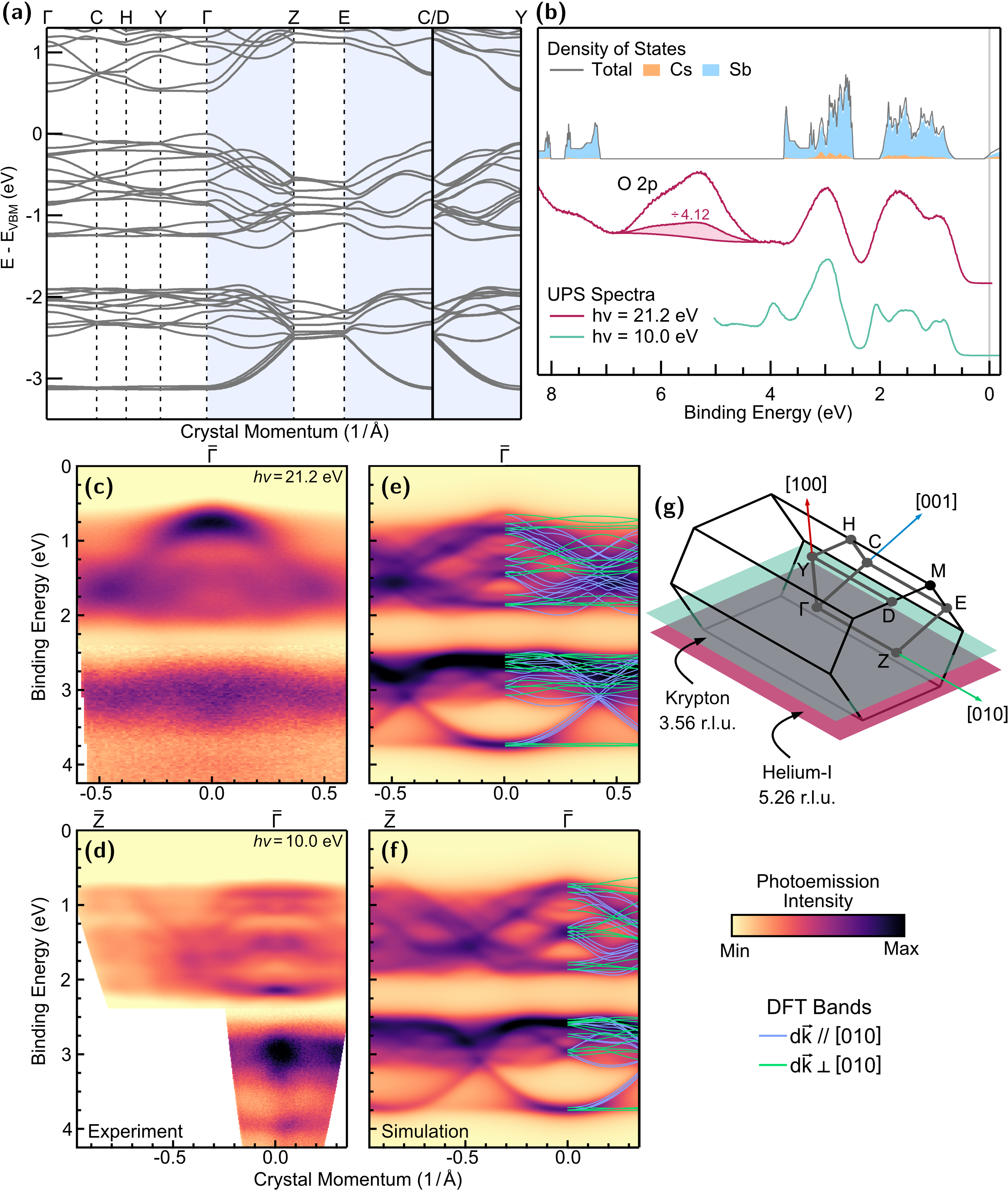}
  \caption{\label{fig:ARPES} Calculated and measured electronic structure of CsSb films grown on SiC (100) (a) DFT calculation of the band structure of monoclinic $\beta$--CsSb. Cuts taken along directions parallel to the [010] axis are highlighed in blue.  (b) Calculated density of states, with Cs and Sb contributions highlighted in orange and blue.  Angle-integrated photoemission spectra illustrating the structure of the near E$_f$ Sb \textit{5p} manifold in green and magenta.  O \textit{2p} states are observed between 4.4 and 6.7 eV; their adjusted weight (divided by the ratio of the O/Sb photoemission cross sections $\sim 4.12)$ is shaded.  Composite angle-resolved spectra taken with Helium-I ($h\nu=21.2$ eV) and Krypton-I ($h\nu=10.0$ eV) light are shown in (c) and (d), respectively. Corresponding simulated spectra for a fiber-texture (100) oriented film are displayed in (e),(f) at out-of-plane momenta of 5.26 and 3.56 r.l.u. (1 r.l.u = $2\pi/d_{(100)}$, with $d_{(100)}\sim 13.92$~\AA). Bands originating from specifically oriented domains are overlaid. (g) Brillouin zone\cite{Hinuma2017,Togo2018} of $\beta$--CsSb with the k-space path from (a) outlined in grey.  Positions, in the folded zone scheme, of the ARPES cuts in (c) and (d) are highlighted in magenta and green, respectively.}
\end{figure*}

We turn now to calculations and measurements of the electronic structure of MBE grown CsSb.  A density functional theory (DFT) calculation of the band structure for bulk-like $\beta-$CsSb is shown in \hyperref[fig:ARPES]{Figure \ref*{fig:ARPES}(a)} and the corresponding calculation for the $\alpha$ phase is included in the supplementary materials.  Given the quasi one-dimensional crystal structure, it is perhaps unsurprising that the resulting electronic structure is also quasi one-dimensional.  Paths parallel to the [010] axis (highlighted in blue) show dispersive features corresponding to hopping along the Sb-Sb chains.  In contrast paths perpendicular to [010], i.e. hopping across a more ionic Sb-Cs-Sb bond, show much flatter dispersion.  Qualitatively, the band structures of the $\alpha$ and $\beta$ phases are quite similar -- the main difference arises from band splittings corresponding to the larger number of inequivalent Sb sites in the monoclinic cell.  The DFT bandwidth of the near $E_f$ manifold, composed primarily of Sb $5p$ states, is calculated to be 3.12 eV in the $\beta$ phase and 3.06 eV in the $\alpha$ phase.  These bandwidths are both substantially larger than those measured for the valence bands of Cs$_3$Sb, 1.2 eV\cite{Epitaxy}, meaning determination of the density of states by ARPES is a good metric for discriminating between the two phases (e.g. see comparison in supplementary materials Fig. S6(b)).  

Indeed, the DFT calculated density of states matches well with \textit{in situ} measurements of the valence band structure, up to an overall shift of $E_f-E_{VBM}=620$~meV, both in terms of peak structure and overall bandwidth, which is measured to be 3.55 eV (see \hyperref[fig:ARPES]{Figure \ref*{fig:ARPES}(b)}).  The position of the Fermi level at 620 meV above the valence band maximum can be attributed to pinning of the chemical potential in the gap.  However, this shift is similar to the DFT calculated gap value of 520 meV -- indicating the Fermi level may lie at or near the the conduction band minimum.  While this suggests native electron doping, no weight is observed at E$_f$ so additional optical and electrical measurements are required to determine the true gap size and the carrier sign.  We note that underestimation of the bandwidth (in this case by $\sim 14$\%) by the electronic structure calculation is similar to previous measurements of Cs$_3$Sb/SiC (100) where the observed bandwidth of the valence states is between 10 and 20\% larger than the LDA prediction.\cite{Epitaxy}  In addition to the expected Sb $5p$ and Cs $6s$ states near the Fermi level, an additional peak is observed between 4.2 and 6.8 eV of binding energy -- associated with the presence of oxygen $2p$ states.  The calculated photoemission cross section for oxygen is enhanced over that of antimony in this energy range by a factor of 4.12,\cite{Yeh1985} so even superficial oxidation of the surface, as suggested by XPS measurements, may result in a large O $2p$ signal in ARPES. A more accurate representation of the oxygen DOS is included in the shaded region of \hyperref[fig:ARPES]{Figure \ref*{fig:ARPES}(b)} where the oxygen weight has been divided by its relative cross section.  Finally, the weak peak observed at 7.5 eV is attributable to Sb $5s$ states which have a diminished cross-section (only 1.7\% of $\sigma_{Sb 5p}$ at $h\nu=21.2$ eV).\cite{Yeh1985}

A surprising consequence of the quasi one-dimensional band structure is that when angle resolved photoemission measurements are performed on these fiber-texture films, momentum resolved features are clearly visible, as is shown in \hyperref[fig:ARPES]{Figures \ref*{fig:ARPES}(c-d)} -- taken with Helium-I ($h\nu = 21.2$ eV) and Krypton-I ($h\nu = 10.0$ eV) light, respectively.  Using the fitted electron affinity, calculated band gap, and measured work function we estimate an inner potential\cite{Damascelli2004} of $V_0\sim 5.74$~eV -- placing the out-of-plane momenta at 5.26 and 3.56 r.l.u. (1 r.l.u = $2\pi/d_{(100)}$, $d_{(100)}=13.92$~\AA) for Helium-I and Krypton-I.  At both photon energies dispersive bands are visible between 1.50-2.25 eV and 4.25-3.50 eV, with a maxima observed at the center of the projected zone, consistent with the DFT calculation.  The fact that the band structure is not completely washed away by the macroscopic rotational disorder is a consequence of the quasi one dimensional nature of the structure: the measured spectrum is an incoherent sum over rotated domains with most of the dispersion arising from domains where [010] is nearly aligned to the electron analyzer slit. Domains of other orientations contribute primarily flat bands and add up to a nearly momentum independent background.

More quantitatively, the spectra can be simulated by calculating a spectral function from the DFT band structure for each rotated domain and then performing a sum over angles.  The results of these calculations, at out-of-plane momenta corresponding to those of the ARPES measurements, are shown in \hyperref[fig:ARPES]{Figure \ref*{fig:ARPES}(e-f)}. This simulation captures many of the salient features of the ARPES spectra even without taking into account optical matrix element effects.  The most well defined features and clearest dispersion are observed near the zone center, with the mismatched lattice constants of the rotated domains blurring the spectra at higher momentum. However, the features from the neighboring zone are still recognizable in the ARPES spectrum \hyperref[fig:ARPES]{Figure \ref*{fig:ARPES}(d)}, as predicted by the simulation in \hyperref[fig:ARPES]{Figure \ref*{fig:ARPES}(f)}.  As expected, the features corresponding to the highly dispersive direction along the Sb-Sb chains (shown in blue) are dominant with a lesser contribution from the flatter bands (shown in green) from paths perpendicular to the chains.

The presence of measurable momentum-resolved features in the ARPES spectra further evidences the high degree of surface order that can be achieved in this system, despite the macroscopic rotational disorder in the films.  Additionally the agreement between the measured spectra and calculated electronic structure from the monoclinic phase further evidences that the phase of the films is predominantly CsSb rather then another member of the Cs-Sb phase diagram.  We note, however, that due to the very similar structure of the Sb-Sb chains in the $\alpha$ and $\beta$ phases the general dispersive features of $\alpha-$ and $\beta-$CsSb are expected to be quite similar.  So, while the measured dispersive features and DOS match well with calculations for $\beta-$CsSb, it is not possible to unequivocally rule out the presence of the $\alpha$ phase with photemission measurements alone.  However, the valence band structure and dispersion observed in these measurements allow Cs$_3$Sb and Sb metal to be confidently excluded as the dominant photoemitting phases of the film.

%%%%%%%%%%%%%%%%%%%%%%%%%%%%%%%% QE & Dosing %%%%%%%%%%%%%%%%%%%%%%%%%%%%%%%%%%%
%%%%%%%%%%%%%%%%%%%%%%%%%%%%%%%%%%%%%%%%%%%%%%%%%%%%%%%%%%%%%%%%%%%%%%%%%%%%%%%%
\section{Quantum Efficiency and Oxidation}
\begin{figure}
  \includegraphics[width=8.5cm]{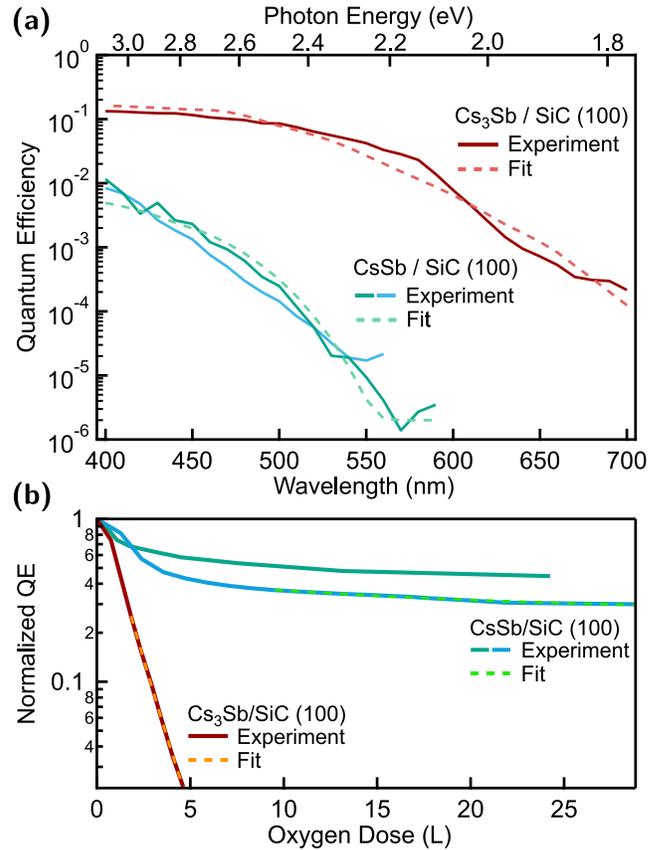}
  \caption{\label{fig:QE} (a) Spectral response of CsSb and Cs$_{3}$Sb thin films grown by codeposition. Fitting of the response curves indicates gives photoemission thresholds of > 580 nm and > 650 nm for CsSb and Cs$_3$Sb, respectively.  (b) The QE of CsSb and Cs$_{3}$Sb as a function of oxygen exposure. Dashed lines represent the best fit to an exponential decay; the ratio between the CsSb and Cs$_3$Sb decay constants shown is 15.}  
\end{figure}

We finish with a discussion of the low energy photoemission properties of CsSb, which are of primary importance to its application as a high brightness photoemitter.  To this end the spectral response of two CsSb films was measured and the results are summarized in \hyperref[fig:QE]{Figure \ref*{fig:QE}(a)}, together with the measurement of a codeposited Cs$_3$Sb reference sample. The films were synthesized on 3C-SiC (100) in a different growth system than those discussed in previous sections, however the structure was again monitored using \textit{in operando} RHEED and the growth temperatures were adjusted to take into account a different thermocouple and heater geometry.  The resulting CsSb films exhibited the same fiber texture RHEED pattern observed in \hyperref[fig:Growth]{Figures \ref*{fig:Growth}(d-e)} and similar QEs of $\sim 10^{-4}$ at 504 nm. Following growth, the samples were transferred in vacuum to a storage chamber (to avoid reaction with residual Cs vapor in the growth chamber), where the spectral response was measured between 700 and 400 nm. The maximum QE observed in this range was approximately $1.2\%$ at 400 nm, which is comparable to Cs$_3$Sb at 590 nm.  The photoemission threshold for each cathode was estimated using the Dowell-Schmerge model\cite{Dowell2009} modified for semiconductors. From this fitting the threshold was estimated to be 2.19 eV for CsSb and 1.65 eV for Cs$_3$Sb.

Following spectral response measurements, the QE degradation of these CsSb photocathodes was measured as a function of oxygen exposure. The samples were exposed to controlled levels of O$_2$ via a leak valve and nozzle with the O$_2$ partial pressure maintained between \smash{5$\times10^{-9}$} and \smash{5$\times10^{-8}$} Torr; the chamber background pressure was below $10^{-9}$ Torr. The QE is reported as a function of nominal oxygen dose (in Langmuir, 1~L equivalent to 1$\times10^{-6}$ Torr $\times$ 1 second) in \hyperref[fig:QE]{Figure \ref*{fig:QE}(b)}. As a control, a high efficiency Cs$_3$Sb cathode was also synthesised in the same growth system and dosed in an identical geometry for a comparable reference.  Laser wavelengths of 400 nm and 532 nm were chosen to measure the QE of the CsSb and Cs$_3$Sb samples, respectively, to make the starting QEs comparable ($\sim 1$\%) for both. Measuring the QE degradation of the two distinct photocathode films starting from percent-level quantum efficiencies demonstrates the critical difference between CsSb and Cs$_3$Sb.  The CsSb films exhibit a resistance to oxidation more than 10 times that of Cs$_3$Sb up to an exposure of 30 L.  Such chemical stability means that the use of CsSb might extend the usable lifetime of alkali antimonide cathodes in photoguns by over an order of magnitude -- extending their use to weeks or months instead of the days long lifetimes of standard alkali antimonide photoemitters. \cite{Luca_rugged}

%%%%%%%%%%%%%%%%%%%%%%%%%%%%%%%% Conclusions %%%%%%%%%%%%%%%%%%%%%%%%%%%%%%%%%%%
%%%%%%%%%%%%%%%%%%%%%%%%%%%%%%%%%%%%%%%%%%%%%%%%%%%%%%%%%%%%%%%%%%%%%%%%%%%%%%%%
\section{Conclusions}
We have demonstrated the synthesis of atomically smooth thin films of CsSb by codeposition of Cs and Sb on both 3C-SiC~(100) and graphene/TiO$_2$~(110). This compound, although less efficient than Cs$_3$Sb, is characterized by $~1\%$ QE at 400~nm, and its photoemission threshold is close to 520~nm; both wavelengths are easily achievable from common laser gain media. This means that it would be equally easy to operate this cathode both at $\sim1$\% QE and near the photoemission threshold, where the lowest emittance is expected. However, the intrinsic emittance has been measured to be minimal at the photoemission threshold only on metal and alkali antimonide photocathodes, \cite{LucaCold} while Cs$_2$Te is a notable exception, \cite{Pierce} so further studies are needed to ascertain its photon energy dependence for CsSb. STM and RHEED studies of the morphology indicate that this compound can be grown atomically smooth, which sidesteps some of the challenges facing the growth of  AA$'_{2}$Sb photoemitters, where physical and chemical roughness limit the realization of the intrinsic emittances of the material.  We have shown that, despite the random in-plane orientation of domains in the film, the surface remains sufficiently ordered to display dispersion in ARPES. Finally, we observe that CsSb has a greatly improved resistance against oxidation over Cs$_3$Sb. This allows preservation of the atomically ordered surface during vacuum suitcase transfers, as revealed by STM and XPS. This robustness would facilitate studies on the intrinsic emittance of ordered high efficiency semiconductors, providing a benchmark for the theoretical study of low energy photoemission on this class of materials.  The superior resistance to oxidation, reasonably high quantum efficiency in the visible range, and exceptionally low surface roughness indicate CsSb is worth considering as a photocathode for future photoinjector beamlines and light sources. 

%%%%%%%%%%%%%%%%%%%%%%%%%%%%%%%%%%%% Methods %%%%%%%%%%%%%%%%%%%%%%%%%%%%%%%%%%%
%%%%%%%%%%%%%%%%%%%%%%%%%%%%%%%%%%%%%%%%%%%%%%%%%%%%%%%%%%%%%%%%%%%%%%%%%%%%%%%%
\section{Methods}
\subsection{Thin Film Synthesis}
Cs$_x$Sb thin films were grown on $10\times 10$~mm 3C-SiC (001) and graphene coated rutile TiO$_2$(110) substrates affixed to custom niobium sample holders in a Veeco Gen10 MBE system (\smash{$P_{base}\sim 3\times10^{-9}$}~torr) at the PARADIM thin film facility (\href{https://www.paradim.org/}{https://www.paradim.org/}).  Substrates were heated using a resistive heater and the temperature monitored via a thermocouple suspended behind the sample holder. Preceding growth, substrates were degassed at 650 $^{\circ}$C for 15 minutes until a clear RHEED pattern was observed and then cooled in vacuum to the deposition temperature. Deposition was performed using molecular beams from an elemental Sb source and a Cs-In alloy source\cite{Dongxue2020,Epitaxy} with fluxes calibrated via quartz crystal microbalance with an accuracy of $\pm 15\%$.  Typical source temperatures were 288-310 C and 407-420 C giving fluxes of 1.6-3.0$\times10^{13}$ and 3.8-4.9$\times10^{12}$ for cesium and antimony, respectively.

\subsection{Sample Characterization}
Following growth, samples were transferred through a UHV manifold (\smash{$P<2\times10^{-9}$}~Torr) to adjacent measurement chambers. The QE was measured using laser diodes, a positively biased collection coil, and a picoammeter. \textit{In situ} ARPES measurements were performed at room temperature in an analysis chamber with base pressure better than \smash{5 $\times$ 10$^{-11}$}~Torr using a Scienta Omicron DA30-L electron analyzer and Fermion Instruments BL1200s plasma discharge lamp generating Helium-I ($h\nu = 21.2$~eV) and Krypton-I ($h\nu = 10.0$~eV) light.  \textit{In situ} x-ray photoelectron spectroscopy (XPS) measurements were performed in the same chamber using a non-monochromated Scienta Omicron DSX400 x-ray source.  Selected samples were transferred, using a UHV suitcase (\smash{$P<5\times10^{-10}$}~Torr) to a separate UHV system for further XPS and STM measurements. X-ray photoelectron spectra were analyzed with an Omicron Sphera II analyzer after excitation by an unmonochromated Mg K$\alpha$ source (Omicron DAR 400). XPS spectra were collected at photoelectron emission angles of $0^\circ$ and $70^\circ$ from the sample's surface normal. In the latter configuration, the reduced photoelectron escape depth (by $\sim 1/3$) enhances the surface sensitivity, making the measurement sensitive to composition gradients and surface contaminants. STM analysis was performed at room temperature in ultrahigh vacuum using a W tip and a Omicron variable-temperature STM. The tunneling conditions were 50-100 pA at –0.5 to –0.8 V.  

\subsection{Spectral Response and Oxygen Dosing}
For the QE measurements and oxidation experiments depicted in Fig. \ref{fig:QE}, sample growth on 3C-SiC(100) substrates was reproduced in a custom built MBE system equipped with \textit{in operando} RHEED and QE measurement capabilities.  The 3C-SiC (100) substrates were annealed at 650$^\circ$C for 1 hour before lowering to a temperature between 160-200$^\circ$C for deposition.  Following growth, the samples were moved to an adjacent UHV chamber (P$\sim 10^{-10}$~Torr) to prevent further reaction with residual alkali metal vapor in the growth system.  There, the spectral response of the samples was measured using an Oriel Apex Monochromator light source, a Newport optical power meter (model 843-R), and an SRS 8340 lock-in amplifier. The photocurrent was collected by biasing a metallic coil placed 5 cm from the sample at +120 V.

For the oxygen dosing experiments, selected samples were returned to the growth chamber, where oxygen was introduced from a leak valve through a nozzle directed at the sample surface; the oxygen partial pressure was measured by a residual gas analyzer. The QE was measured at a single wavelength, provided by a laser diode, and the photocurrent measured by monitoring the drain current from the electrically floating sample holder (biased at $-40$~V). Dosing experiments were performed on both CsSb and Cs$_3$Sb samples in the same configuration to rule out differences between the measured pressure at the gauge and the pressure at the sample surface arising to the chamber's pumping layout.
The QE versus oxygen dose curves have been fitted with a simple exponential decay. In particular for the CsSb data, the initial faster decay within the first 10L ,visible in \hyperref[fig:QE]{Figure \ref*{fig:QE}~(b)}, was disregarded. The ratio between the decay constants of CsSb and Cs$_3$Sb is found to be $\sim 7-15$.

\subsection{Density Functional Theory Calculations}
Plane-wave density functional theory calculations of $\alpha$ and $\beta$-CsSb were performed using GGA-PBE exchange-correlation functionals\cite{Perdew1996} and SG15 norm-conserving pseudopotentials\cite{Schlipf2015a} implemented in JDFTx\cite{Sundararaman2017}.  For the monoclinic ($\beta$-CsSb) structure a plane-wave cutoff of 40 Hartrees was used and optimized structural parameters of $a=15.50$~\AA, $b=7.50$~\AA, $c=14.55$~\AA, and $\beta=113.82^{\circ}$ were obtained from a relaxation calculation; calculations of total ground state energy and DOS used a mesh of $5\times 7\times 5$.  To enable efficient interpolation of the electronic band structures to arbitrary crystal momenta and for more accurate calculation of the DOS, the Wannier interpolation technique\cite{Marzari2012} was used to generate a maximally localized Wannier basis set\cite{Marzari1997} using a supercell of $4\times 7\times 4$ primitive cells using linear combinations of bulk Bloch bands at binding energies from 0 to 11 eV below the valence band maximum. To generate the simulated ARPES spectra in Fig. \ref{fig:ARPES}(e-f) a set of spectral functions $A(\textbf{k},\omega)\sim 1/((\omega-\epsilon_{i,\textbf{k}})^2+\Sigma''^2)$ was then generated from the wannier interpolated eigenvalues, $\epsilon_{i,\textbf{k}}$, and an imaginary self energy of $\Sigma''=75$~meV.

%%%%%%%%%%%%%%%%%%%%%%%%%%%%%%%%% Ending Matter %%%%%%%%%%%%%%%%%%%%%%%%%%%%%%%%
%%%%%%%%%%%%%%%%%%%%%%%%%%%%%%%%%%%%%%%%%%%%%%%%%%%%%%%%%%%%%%%%%%%%%%%%%%%%%%%%
\section*{Supplementary Material}
See the supplementary material for additional RHEED and STM images, further display and analysis of XPS data, and a discussion of the crystal and electronic structure of the orthorhombic phase $\alpha$-CsSb.
\section*{Acknowledgments}
This work was supported by the U.S. National Science Foundation Grant PHY-1549132, the Center for Bright Beams, and by the National Science Foundation (Platform for the Accelerated Realization, Analysis, and Discovery of Interface Materials (PARADIM)) under Cooperative Agreement No. DMR-2039380.  Work by C.T.P. and K.M.S. also acknowledges support from NSF DMR-2104427 and AFOSR FA9550-21-1-0168. J.M.M. acknowledges support from DOE grants DE-AC02-76SF00515 and DE-SC0020144. This work made use of the Cornell Center for Materials Research Facilities supported by the National Science Foundation under Award Number DMR-1719875. Substrate preparation was performed in part at the Cornell NanoScale Facility, a member of the National Nanotechnology Coordinated Infrastructure (NNCI), which is supported by the National Science Foundation (Grant NNCI-2025233). The authors thank Sean C. Palmer for his assistance in substrate preparation.
\section*{Author Declarations}
\subsection*{Conflict of Interest}
The authors have no conflicts to disclose
\subsection*{Author Contributions}
Sample synthesis was performed by A.G., C.T.P., C.A.P., E.E., and W.J.D. under the supervision of J.M.M., K.M.S, and D.G.S with assistance from H.P..  QE experiments were performed by A.G., C.A.P., E.E., and C.T.P..  C.T.P. and B.D.F. performed the \textit{In situ} ARPES and XPS measurements with assistance from L.M. and C.H. and data was analyzed by C.T.P. and V.A.. STM and angle dependent XPS measurements were performed by J.B. and W.J.D., and A.G. collaborated on the XPS data analysis under the supervision of M.A.H..  DFT calculations were performed by C.T.P. and J.K.N. under the supervision of K.M.S. and T.A.A.  A.G., C.T.P., C.A.P. and J.M.M. prepared the manuscript with input from all authors.
\section*{Data Availability}
The data that support the findings of this study are available within the paper and supplementary material. Additional data related to the growth and structural characterization are available at (DOI pending publication). Additional data connected to the study are available from the corresponding author upon reasonable request.
\section*{References}
%\nocite{*} % Include all references
\bibliography{AA_CsSb_Bibliography}

\end{document}